# Mode-Specific Dynamics of $CO_2$ Hydrogenation on Copper: The Hidden Role of Molecular Rotation

Junfan Xia[1], Zhikai Jiang[1], Yaolong Zhang[1], Bo Peng[2], Hua Guo[3], and Bin Jiang[1,4*]

1. State Key Laboratory of Precision and Intelligent Chemistry, Department of Chemical Physics, University of Science and Technology of China, Hefei, Anhui 230026, China
2. State Key Laboratory of Petroleum Molecular and Process Engineering, SINOPEC Research Institute of Petroleum Processing Co., Ltd., Beijing 100083, China.
3. Department of Chemistry and Chemical Biology, Center for Computational Chemistry, University of New Mexico, Albuquerque, New Mexico 87131, USA
4. Hefei National Laboratory, University of Science and Technology of China, Hefei, 230088, China

*: Corresponding author: bjiangch@ustc.edu.cn

## Abstract

Catalytic hydrogenation of $CO_2$ to formate on copper is a key elementary step for $CO_2$ utilization. Previous experimental and theoretical studies suggested an Eley-Rideal mechanism for this reaction, promoted by bending vibrational excitation, yet direct state-resolved evidence remains lacking. Here, we present first-principles dynamical predictions for $CO_2$ hydrogenation on Cu(111) based on an accurate full-dimensional neural network potential energy surface. Our calculations near-quantitatively reproduce the measured reaction probabilities, including their nozzle-temperature and incidence-energy dependence. Our state-resolved results indicate that while vibrational excitation of the bending mode enhances reactivity, it alone cannot account for the observed reactivity increase with nozzle temperature. Instead, rotational excitation plays a dominant role, mainly attributable to the significant change in anisotropy of the molecular polar orientation as $CO_2$ accesses the transition state. This mode-specific insight reinforces the hidden role of rotation in surface reactivity, opening new avenues for state-selective control of $CO_2$ hydrogenation on heterogenous catalysts.

## Introduction

The rapidly escalating atmospheric concentration of $CO_2$ poses a challenge of global climate change. Utilizing $CO_2$ as a feedstock and renewable hydrogen, catalytic hydrogenation of $CO_2$ to value-added chemicals and fuels represents a compelling strategy toward carbon neutrality and sustainable development[1]. Formate (HCOO) is a pivotal intermediate in various $CO_2$ hydrogenation processes, acting as a branching point toward a spectrum of valuable chemicals[2], such as formic acid[3], methanol[4], and other hydrocarbons[5] . A precise understanding of formate formation as a key elementary step of $CO_2$ hydrogenation, particularly on industrially relevant copper-based catalysts[6], is therefore essential for elucidating the mechanisms of relevant catalytic processes and for guiding the design of more effective catalysts for $CO_2$ conversion.

Earlier experiments and density functional theory (DFT) calculations suggested that formate synthesis ($CO_2 + 1/2H_2 \rightarrow HCO_2^*$) on copper surfaces proceeds mainly via an Eley-Rideal (ER) mechanism[7-9], where $CO_2$ reacts directly with adsorbed $H^*$ species on the surface. This mechanism, which is supported by the experimentally observed pressure dependence[7] and the kinetic insensitivity to surface structure[8, 9], implies that gaseous $CO_2$ does not equilibrate to the surface before the association reaction. On the other hand, formate decomposition on copper catalysts—the reverse of formate synthesis—was found to produce non-thermal $CO_2$ molecules whose mean translational energy is independent of surface temperature ($T_s$)[10]. This non-equilibrium behavior has been confirmed by dynamical calculations[11, 12], lending further support to the proposed ER mechanism for formate synthesis.

More recently, Quan *et al.* reported the first direct experimental investigation of the ER reaction dynamics in formate formation from $CO_2$ and pre-adsorbed H on Cu surfaces[13]. The measured initial reaction probability ($P_0$) is largely independent of surface temperature and structure, and elevated by both translational and vibrational excitation. Particularly, the substantial increase of $P_0$ with nozzle temperature ($T_n$) suggests that the vibrational energy of $CO_2$ is significantly more efficient than its translational energy in driving this association reaction. This observation was qualitatively supported by DFT calculations, which identified a bent transition state (TS) for the ER reaction,[13] although confirmation is needed through more precise initial state-resolved measurements. However, developing efficient schemes for selectively exciting vibrational states of $CO_2$ remains challenging. Therefore, accurate quantum-state-resolved theoretical investigations[14] are highly desirable alternatives for achieving a complete, mode-specific understanding of $CO_2$ hydrogenation to formate.

In this Letter, we address this need by reporting accurate predictions of the reaction probability of gaseous $CO_2$ and surface H atoms on Cu(111) based on the first full-dimensional neural network potential energy surface (PES) for this system trained on extensive DFT data. Quasi-classical trajectory (QCT) simulations with this PES efficiently predict both thermally averaged and state-resolved $P_0$ as low as $10^{-5}$, achieving near-quantitative agreement with all available experimental data. Furthermore, the calculated results reveal strong mode-specific vibrational enhancement in this association reaction, with the bending mode exhibiting the largest vibrational efficacy. However, this alone does not quantitatively explain the substantial increase in reactivity observed with rising nozzle temperature in earlier experiments[13]. Instead, we find that rotational excitation of the impinging $CO_2$ plays a much more dominant role in reactivity enhancement, which can be attributed to a narrow reaction cone of acceptance arising from the significant increase in PES anisotropy as $CO_2$ approaches the surface H atom. Only when rotational excitation is included can the experimental observations be reproduced quantitatively.

## Results

### Reaction pathway and potential energy surface

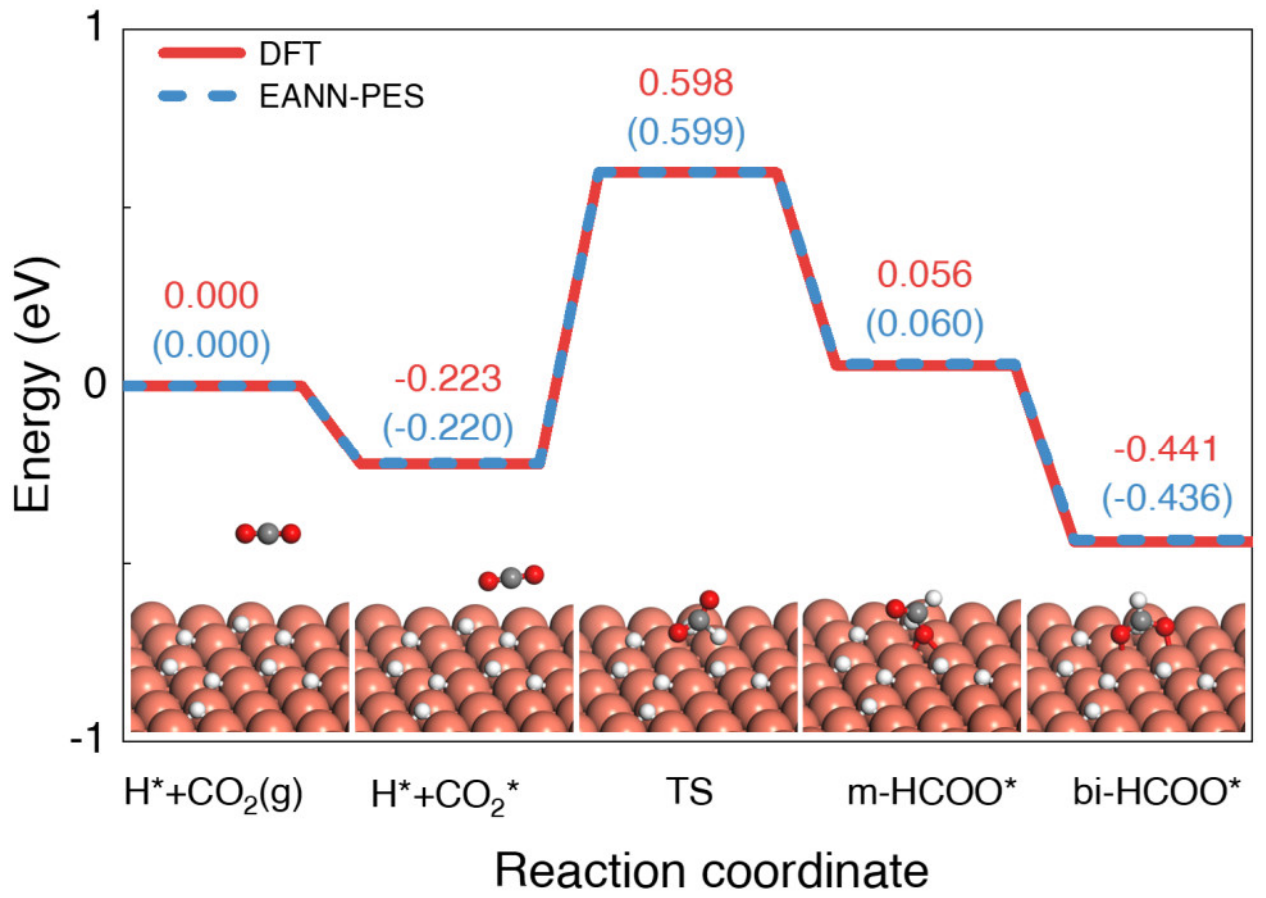

**Figure 1. Reaction energy profiles for formate formation.** Comparison of stationary-point energies along the minimum energy path of the ER-mediated $CO_2$ hydrogenation on Cu(111), optimized with DFT (solid line) and with the EANN PES (dash line).

Briefly, the PES was constructed by means of the Embedded Atom Neural Network (EANN) approach[15]. The training data were generated from DFT calculations performed with the Vienna Ab initio Simulation Package (VASP)[16] using the optPBE-vdW functional[17]. A four-layer slab model consisting of a $4 \times 4$ unit cell of Cu(111) covered by eight H atoms was used, corresponding to an H coverage of 0.5 ML in the experiment[13]. The top three layers of Cu atoms and H atoms were free to move, while the bottom layer was fixed. More details on the PES construction and DFT calculations are given in Supporting Information (SI). Figure 1 compares the DFT and PES minimum energy paths (MEPs) for formate formation between $CO_2$ and $H^*$ on Cu(111) surface. The PES reproduces precisely the DFT energy profile, as well as the geometries and harmonic frequencies of stationary points (see Table S1 and Figure S1 in SI). The $CO_2$ physisorption energy predicted by the PES is 0.220 eV, aligning well with the experimental estimate (0.248 eV)[18]. The TS for the association of $CO_2$ and $H^*$ to form monodentate formate (m-HCOO$^*$) features a bent $CO_2$ geometry with the O-C-O angle of 144.5° and a barrier height of 0.599 eV. The transformation from m-HCOO$^*$ to bidentate formate (bi-HCOO$^*$) is nearly barrierless with an exothermicity of 0.496 eV.

**Reaction probabilities**

QCT calculations were first performed with the PES to obtain thermally-averaged reaction probabilities of $CO_2$ association with $H^*$ on Cu(111). To enable a fair comparison with the experiment, the initial vibrational and rotational state populations of $CO_2$ were sampled from the Boltzmann distributions according to the experimental conditions. In particular, the vibrational temperature was taken as the nozzle temperature ($T_n$); while the rotational temperature ($T_r$) was not accurately measured experimentally and is generally considered to be effectively cooled to a small fraction of $T_n$—taken as 5% of $T_n$ in this work. The initial surface configurations were also sampled by equilibrated molecular dynamics according to the surface temperature ($T_s$).[13] A recently proposed adsorbate Gaussian Binning (AGB) scheme[19] was applied in the product state analysis to minimize the possible influence of zero-point energy leakage.

Figure 2 compares the experimental and theoretical $P_0$ under a diverse set of experimental conditions. At a fixed translational energy of $CO_2$ along surface normal ($E_t$=1.97 eV) with $T_n$=1000 K, corresponding to a thermally-averaged vibrational energy of $\overline{E}_v$=142 meV, our QCT results reproduce well the experimentally observed $T_s$-independence of $P_0$ in the range of 120–220 K. This is clearly suggestive of an ER mechanism, in which $CO_2$ is not equilibrated with the Cu(111) surface but directly reacts with the pre-adsorbed H atoms.

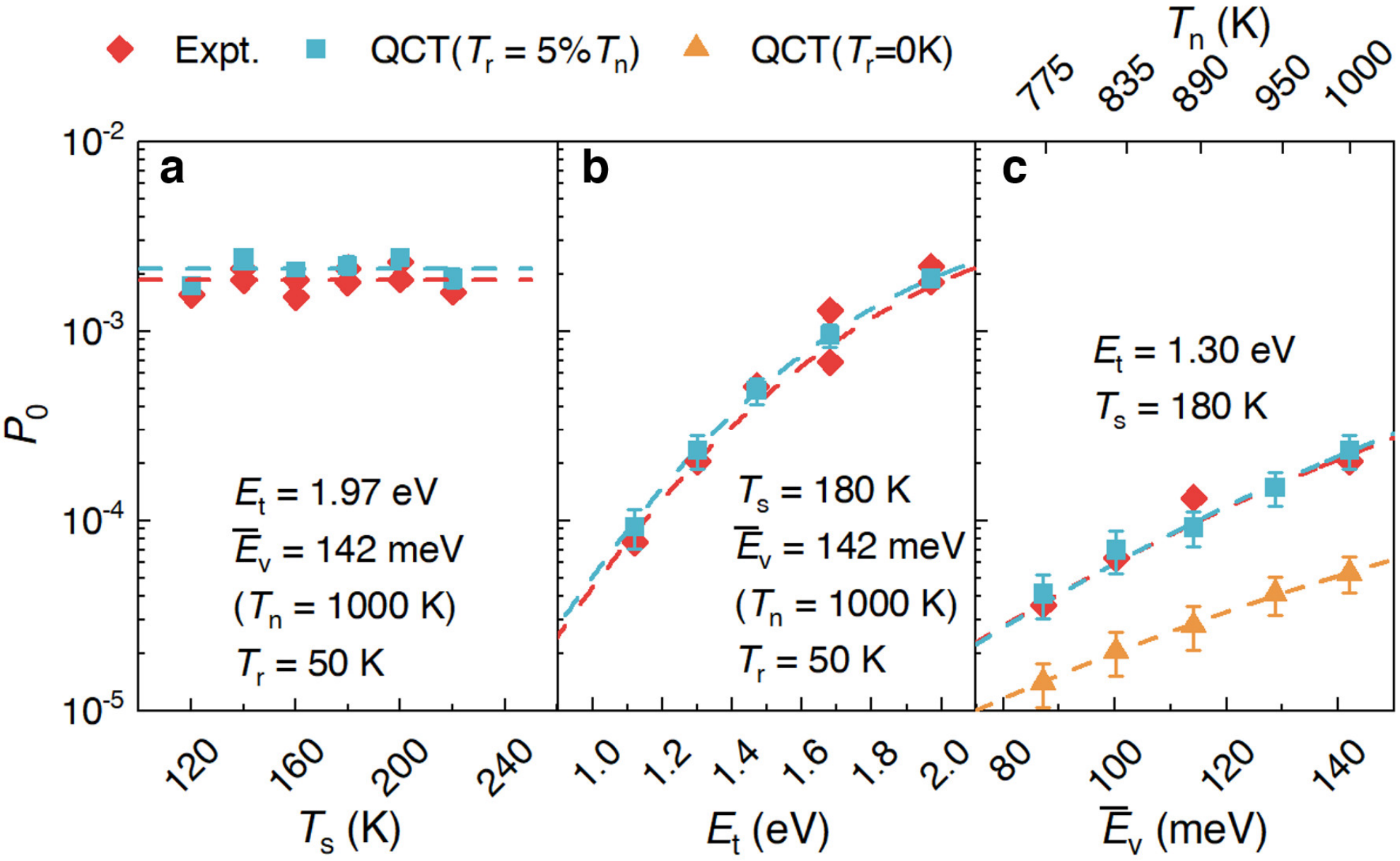

**Figure 2. Comparison of the calculated and experimental reaction probabilities.** $P_0$ of formate formation by $CO_2$ and H-covered Cu(111), as a function of (**a**) $T_s$ with $E_t$=1.97 eV and $T_n$=1000 K ($\overline{E}_v$=142 meV), (**b**) $E_t$ with $T_n$=1000 K and $T_s$=180 K, and (**c**) $\overline{E}_v$ (or $T_n$) with $E_t$=1.30 eV and $T_s$=180 K ($T_r$=5%$T_n$ or 0 K). The pre-adsorbed hydrogen atoms were prepared at a saturated coverage of 0.5 ML. Dashed lines are fitted S-shaped reactivity curves (detailed in the SI). The statistical uncertainties of the calculated results are indicated in the figure by error bars.

Figures 2(b) and (c) show the influence of $E_t$ with $\bar{E}_v$ fixed at 142 meV and of $\bar{E}_v$ with $E_t$ fixed at 1.30 eV on the reaction probability, respectively. Our QCT results find near-quantitative agreement with experimental data, demonstrating that both $E_t$ and $\bar{E}_v$ can significantly enhance the reactivity. Following the experimental definition[13], the effectiveness of vibration relative to translation—tentatively neglecting rotational effects at different $T_n$—was approximately quantified by the mean vibrational efficacy, defined as $\bar{\eta}_v = \Delta E_t / \Delta\bar{E}_v$, where $\Delta E_t$ and $\Delta\bar{E}_v$ are the differences in $E_t$ and $\bar{E}_v$ relative to their reference energies ($E_t^{ref}$=1.30 eV and $\bar{E}_v^{ref}$ =142 meV) required to reach a given $P_0$,[13] respectively. A $\bar{\eta}_v$ value greater than unity indicates that vibrational excitation on average promotes the reaction more effectively than translational excitation, and vice versa. The mean vibrational efficacy estimated from QCT data in Figures 2b and 2c is about 5.8–7.4, in reasonably good agreement with the experimental estimate of 7.9.[13] This value is considerably larger than those state-resolved vibrational efficacies previously reported in dissociative chemisorption experiments of $CH_4$ ($\bar{\eta}_v$ = 0.5–1.4)[20, 21] and $D_2O$ ($\bar{\eta}_v$ =1.1)[22] on Ni(111), or predicted by theories for $CO_2$ dissociation on Ni(100) ($\bar{\eta}_v$=0.06–0.37[23] or 1.3–2.2[24]) and Cu(110) ($\bar{\eta}_v$=0.7–2.3[25] or 1.2-1.6[26]), depending on the theoretical method and incidence energy. The surprisingly high mean vibrational efficacy in this bond-forming process was attributed to thermally populated excited states of the bending modes[13], as DFT analysis shows that the $CO_2$ geometry bends significantly at the TS.

However, this approximate assessment of vibrational efficacy based on the mean vibrational energy for a given $T_n$ unavoidably includes rotational contributions, because the rotational state population depends on $T_r$, which itself varies with $T_n$ in our simulations. Indeed, the influence of the rotational excitation on $P_0$ was not discussed in the experiment[13]. To extract the rotational effect, we performed additional QCT calculations assuming complete rotational cooling in the nozzle by setting $T_r$=0 K. The comparison of QCT results of $T_r$=0 K and $T_r$ = 5%$T_n$ in Figure 2c shows that, for each $T_n$ and $E_t$, even a low rotational temperature — corresponding to mild rotational excitation of $CO_2$ — can enhance the reactivity several times. Analogous to vibrational efficacy, we can roughly estimate rotational efficacy relative to translation from these data. For a given $P_0$, the rotational efficacy relative to vibration can be first estimated from the two curves in Figure 2c, yielding values in the range of 11.9–19.0. When combined with the large mean vibrational efficacy predicted earlier, this yields a mean rotational efficacy relative to translation as high as 75.0–140.6. Such an unprecedentedly large rotational efficacy underscores a stronger influence of $CO_2$ rotation than vibration, representing an unexpected new dimension to facilitate this ER reaction.

**Effects of vibrational and rotational excitations**

To isolate and better quantify the respective effects of vibrational and rotational excitation of $CO_2$ on this ER reaction, we further performed quantum state-specific QCT calculations (details of initial sampling are given in SI). In Figure 3, the initial state-selected reaction probabilities for several low-lying states of $CO_2$ are compared as a function of translational energy. The fundamental frequencies of the symmetric stretching ($\nu_1$), bending ($\nu_2$), and antisymmetric stretching ($\nu_3$) modes of $CO_2$ are 1305.7, 622.1, and 2330.1 $cm^{-1}$ on the PES, respectively. Accordingly, a state-specific vibrational efficacy can be clearly defined by the ratio of translational energy difference versus each vibrational excitation energy for a given $P_0$, *i.e.*, $\eta_\nu = \Delta E_t / \Delta E_v$. All vibrationally excited states of $CO_2$ are found to apparently enhance the reactivity between $CO_2$ and surface H atoms, compared to the ground state of $CO_2$. However, their relative efficacy is strongly mode specific. Vibrational efficacies of the symmetric ($\eta_{\nu_1}$=1.3–1.8) and antisymmetric stretch ($\eta_{\nu_3}$=0.7–0.9) modes are relatively small. In comparison, exciting either of the doubly degenerate bending modes exhibits the strongest vibrational efficacy, $\eta_{\nu_2}$=3.0–3.9. Yet, it is still considerably lower than the experimental $\bar{\eta}_\nu$ value of 7.9 derived from the thermally-averaged vibrational energy with varying $T_n$.[13]. Obviously, the two stretching-mode excited states are expected to contribute negligibly to the reactivity due to their low populations under thermal conditions and their insignificant vibrational efficacies. While the increased reactivity at higher $T_n$ may be partially attributable to the higher reactivity of bending-excited states, this contribution alone cannot quantitatively account for the large enhancement observed experimentally.

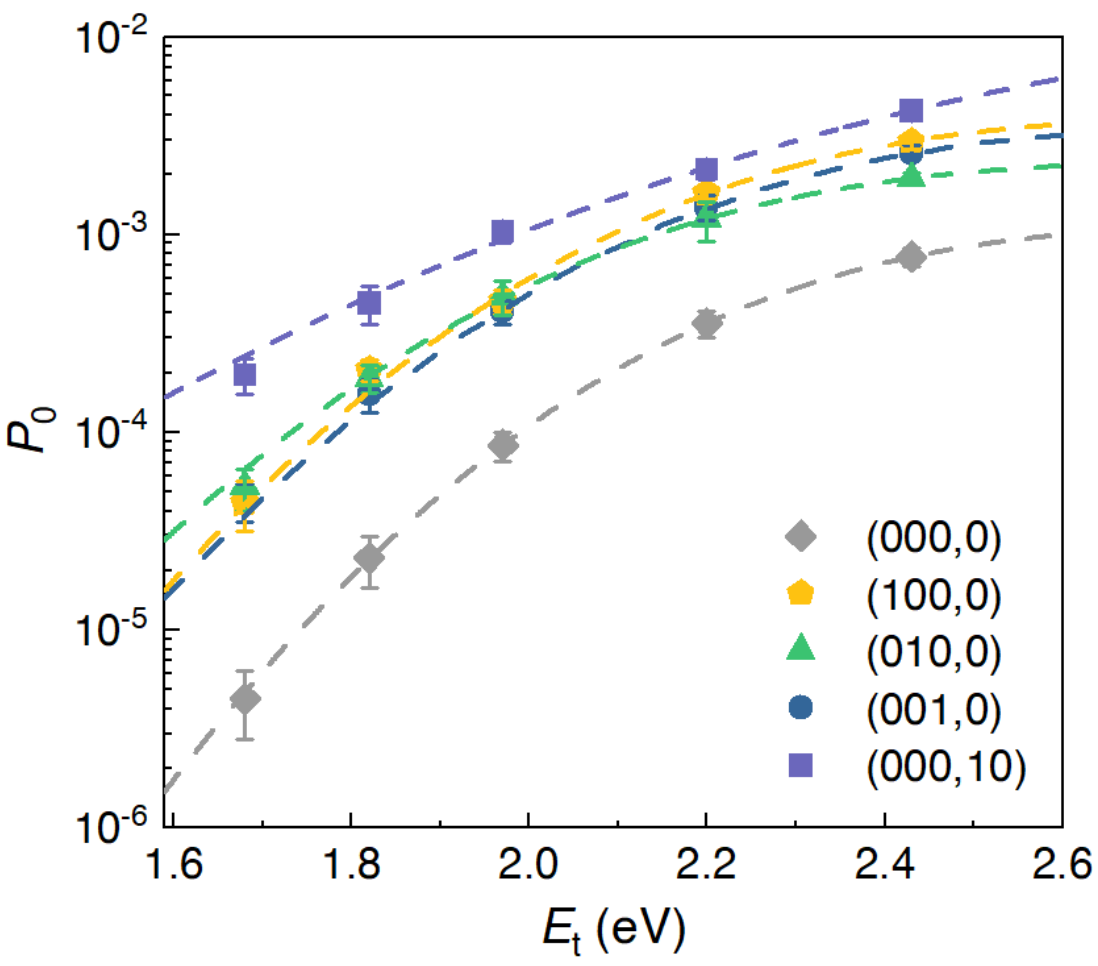

**Figure 3. State-resolved reaction probabilities.** Calculated state-selected reaction probabilities $P_0$ as a function of translational energy for different (rotationless) vibrational states of $CO_2$, as well as for a rotationally excited state ($J$=10) in the vibrationally ground state, with its rotational energy comparable to the mean rotational energy at $T_r$=50 K. Each state is labelled by $(n_1 n_2 n_3, J)$, where $n_1$, $n_2$, and $n_3$ are quantum numbers for the symmetric stretching, bending, and antisymmetric stretching vibrational modes, respectively, and $J$ denotes the rotational quantum number.

Also shown in Figure 3 is the reaction probability curve of a rotationally excited state of $CO_2$ (000, $J$=10), whose rotational energy is comparable to the mean rotational energy at $T_r$=50 K. This moderate rotational excitation leads to a remarkable enhancement in reactivity compared to the ground state and the rotational efficacy reaches a striking range of 90.9–150.9. This state-specific result confirms the mean rotational efficacy indirectly estimated above by differentiating zero-$T_r$ results from thermally averaged ones at $T_r$=5%$T_n$. Such an exceptionally high rotational effect has rarely been seen in surface reactions, particularly for a symmetric molecule. Indeed, previous studies on the dissociation of symmetric molecules like $H_2$, $H_2O$, and $CH_4$ have revealed only weak rotational effects.[27-31] Rotational excitation is found to be more efficient in activating HCl dissociation on metal surfaces[32-34], although the reported high rotational efficacies are limited to the high energy and high-$P_0$ regime, which decrease sharply with reactivity and the degree of rotational excitation. In contrast, the presently predicted rotational efficacy for $CO_2$ hydrogenation is remarkable in the low-$P_0$ regime and at low rotational temperatures, making it more relevant to $CO_2$-related catalytic processes under realistic conditions. As have discussed by Quan *et al.*[13], the extrapolated reaction probability from their experimental data of supersonic beams is comparable to that measured in high-pressure bulb experiments for formate synthesis[9, 35] at a reaction temperature of 353 K. Given its remarkable efficacy, the rotational energy of $CO_2$ likely dominates the reaction kinetics in formate formation under industrial reaction conditions, where gaseous reactants near the catalysts should become sufficiently heated in reactors.

## Discussion

Since this direct ER reaction occurs rapidly, the vibrational mode specificity can be readily rationalized by the Sudden Vector Projection (SVP) model[36], which links the efficacy of a molecular mode in promoting the reaction to the overlap between the normal mode vector and the reaction coordinate. Practically, ensuring that the reactant configuration aligns maximally with the TS structure, the eigenvector of a specific reactant vibrational mode is projected onto the reaction coordinate, *i.e.*, the imaginary-frequency mode at the TS. Since the TS features a bent O-C-O angle ($\theta$) of

~145°, as shown in Figure 4a, the degenerate bending modes of $CO_2$ exhibit large overlaps with the reaction coordinate, yielding an average SVP value of 0.46, considerably higher than that of translation (0.23). In contrast, symmetric and antisymmetric stretching modes of $CO_2$ yield much lower SVP values of 0.014 and 0.093, respectively. This analysis indicates that the bending vibration is the most effective vibrational mode in promoting formate formation, consistent with the conclusion drawn by Quan *et al.* based on the TS structural analysis.[13] The SVP model can also predict the rotational enhancement of reactivity[37]. However, since the reactant rotation can be defined arbitrarily in three-dimensional space, we adopt a well-defined set of axes related to the reaction to avoid ambiguity. As illustrated in Figures 4b-4d using the TS structure as a reference, we define rotations of the $CO_2$ moiety about three orthogonal axes through the carbon atom: the axis perpendicular to the $HCOO^*$ plane (the principal *a* axis), the C–H axis along the relative translation (*b* axis), and the axis perpendicular to both (*c* axis), *i.e.* the molecular axis when $CO_2$ is linear. The corresponding SVP values are 0.14, 0.054, 0.00, respectively, indicating that the cartwheel-like rotation about the principal axis is well coupled with the reaction coordinate and plays an important role for promoting the reaction.

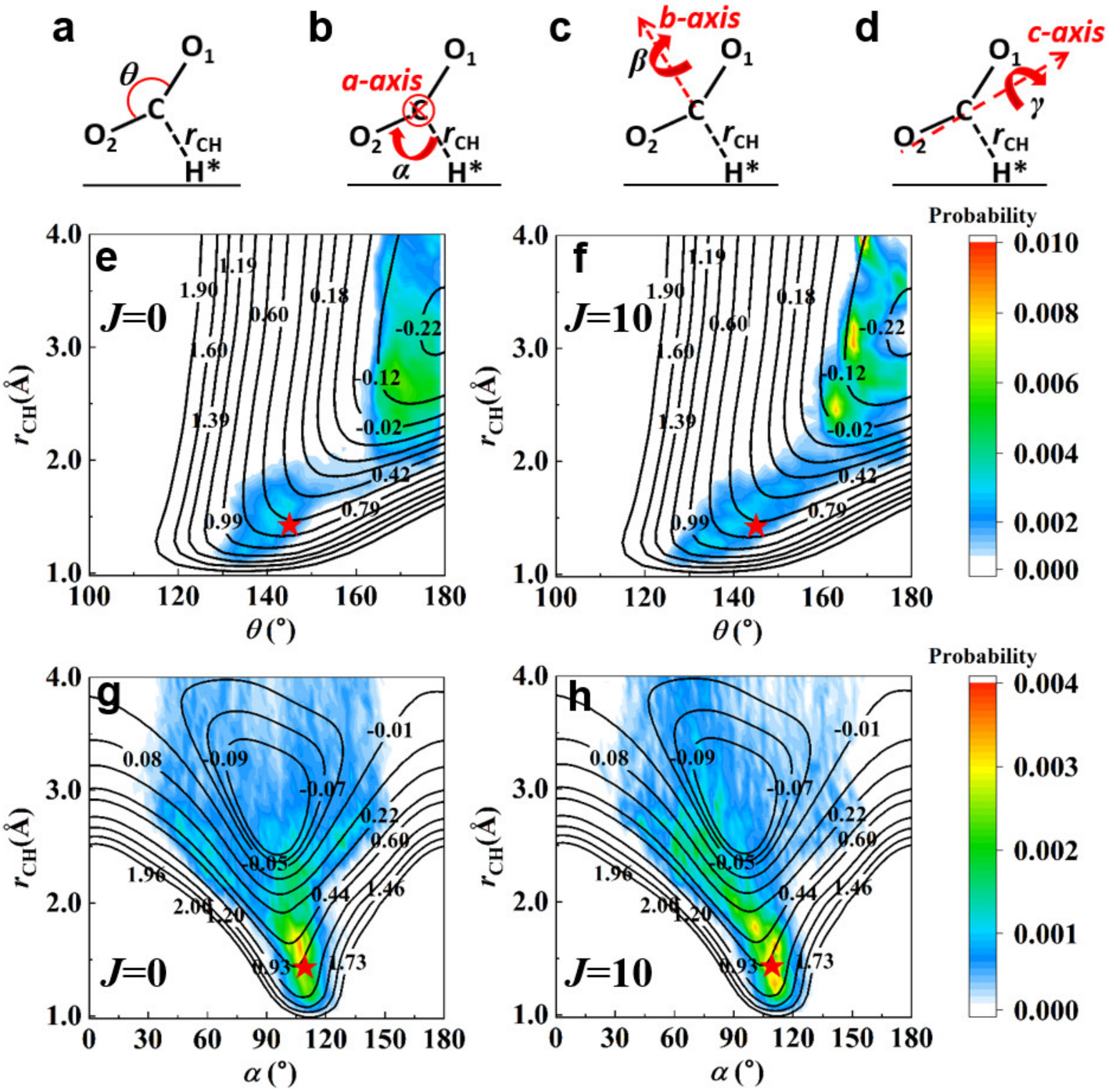

**Figure 4. Two-dimensional PES cuts and projected probability density distributions of reactive trajectories.** Schematic geometries showing the bending (**a**) and rotational angles (**b-d**) of $CO_2$ about the three axes defined in the text. Contours of the PES are shown as a function of: (**e-f**) the C-H distance ($r_{CH}$) and the O-C-O angle ($\theta$); (**g-h**) the C-H distance ($r_{CH}$) and the orientational angle ($\alpha$) of $CO_2$ relative to the C-H vector, rotating about the principal axis defined in panel **b**. The spatial probability density distributions of reactive trajectories for the initial rotational states of $J$=0 (**e** and **g**) and $J$=10 (**f** and **h**) are projected onto the PES cuts to illustrate how trajectories access the relevant regions of configuration space. The H positions and the molecular

approaching direction are kept at their values of the TS. The TS location is marked by red stars in the contour plots.

The origin of the extraordinary rotational effect is better visualized through the comparison of the spatial probability density distributions of reactive trajectories for the initial rotational states of $J$=0 and $J$=10, projected onto the two-dimensional PES contours in Figure 4. Here, Figures 4(e) and 4(f) show that the $CO_2$ molecule must bend from a linear geometry to access the TS with an O-C-O angle ($\theta$) of ~145°, as discussed above. Interestingly, reactive trajectories of initially rotating molecules exhibit a more extended $\theta$-distribution and a higher density in bent geometries before reaching the TS than non-rotating ones. This reveals that the intrinsic coupling between rotation and bending vibration facilitates energy transfer from initial rotation into vibrational excitation in bending-modes, which partially explains why the SVP values for the rotational degrees of freedom alone is not particularly revealing. This behavior is further supported by product vibrational state distributions of scattered $CO_2$ molecules (Table S2 in SI), where initially rotationally excited molecules lead to higher populations of bending-mode excited products than their non-rotating counterparts.

More interestingly, Figures 4(g) and 4(h) display that the PES features increasingly strong anisotropy with respect to the orientational angle ($\alpha$) of $CO_2$ relative to the C-H axis, corresponding to a cartwheel-like rotation about the principal axis defined above, along with an apparent change in this orientational dependence, as $CO_2$ approaches the surface H atom. Accordingly, the incoming $CO_2$ may first steer by the attraction toward a shallow adsorption well ($\alpha \approx 91^{\circ}$). However, it must reorient toward a different and narrow angular window to overcome the TS ($\alpha \approx 108^{\circ}$) for reaction. In this context, an initial rotational angular momentum can assist a molecule that has steered toward the opposite orientation to reorient toward the tight entrance to the TS. This is clearly seen in the $J$=10 trajectories, where a large fraction of molecules are first over-steered into the region of small $\alpha$ values—deviating from the MEP—and then reorient back to the favorable orientation of the TS. In contrast, fewer initially non-rotating $CO_2$ molecules that undergo such orientational change ultimately react, resulting in lower reactivity compared to rotating $CO_2$ molecules. It is worth noting that the PES with respect to rotations about the other two orthogonal axes is much more isotropic (Figure S2 in SI), indicating that corresponding angular momenta about are less relevant to the reaction. This finding predicts a potentially strong steric effect, where the cartwheel-like rotation is likely more effective than the helicopter-like rotation in promoting this ER reaction, awaiting further experimental verifications. This feature of the PES can be likened to an angular "cone of acceptance"[38] for this reaction, which requires some cartwheel rotational angular momentum to enter. In this regard, the pronounced rotational effects in surface reactivity may not be limited to the reaction studied here, as the ER mechanism has been identified in $CO_2$ hydrogenation processes on various catalysts[39, 40], often involving a significant change in the polar orientation of $CO_2$ along the reaction pathway. Further dynamical investigations on these processes are desirable to explore this possibility.

## Conclusion

To conclude, we report a full-dimensional first-principles PES for the ER reaction of $CO_2$ on a H-saturated Cu(111) surface, based on which QCT simulations yield near-quantitative agreement with experimental reaction probabilities at various surface and nozzle temperatures, and incidence energies. We find that the observed enhancement of reactivity with increasing nozzle temperature in experiment is not simply due to initial vibrational excitation, but is instead largely convoluted with a much stronger rotational effect that was completely overlooked previously. More detailed state-resolved QCT results quantitatively identify the vibrational and rotational mode specificity in this reaction. The bending modes of $CO_2$ exhibit strongest vibrational efficacy among vibrational modes, but rotational efficacies are even 30–40 times higher. The SVP analysis shows a large overlap between the bending mode and the reaction coordinate, explaining its promotion on reactivity. The remarkable rotational enhancement is partially due to the intrinsic coupling between rotation and bending motion, and is more strongly correlated with the non-monotonic change in anisotropy along the MEP. While non-rotating molecules may be steered away from the TS by the anisotropy in the entrance channel, cartwheel-like rotation can help the molecule reorient to the favorable orientation of the TS. These findings not only provide unprecedentedly detailed insights into this polyatomic ER reaction, but also carry more broad and practical implications on realistic catalytic processes of $CO_2$ hydrogenation via an ER mechanism. The critical role of rotation in bond formation on catalytic surfaces opens a new way of controlling state-resolved catalytic chemistry at gas-surface interfaces.

## Code availability

The EANN code is available via Github at https://github.com/bjiangch/EANN. The heavily modified VENUS program is available from the corresponding author on request.

## Acknowledgements

This work was supported by Quantum Science and Technology-National Science and Technology Major Project (2021ZD0303301), National Natural Science Foundation of China (22325304, 22221003, and 92570205), an R&D contract from China Petroleum & Chemical Corp. (Grant 122003), and the National Advanced Talent Cultivation Center for Chemistry, USTC. Molecular dynamics simulations and the PES model training were performed on the robotic AI-Scientist platform of Chinese Academy of Sciences, the Supercomputing Center of USTC, and Hefei Advanced Computing Center. We thank Prof. Jiamei Quan for helpful discussion on their experiments.

## Competing interests

The authors declare no competing interests.

# References

(1) Ye, J.; Dimitratos, N.; Rossi, L. M.; Thonemann, N.; Beale, A. M.; Wojcieszak, R. Hydrogenation of CO2 for sustainable fuel and chemical production. *Science* **2025**, *387* (6737), eadn9388.
(2) Ye, R.; Ding, J.; Reina, T. R.; Duyar, M. S.; Li, H.; Luo, W.; Zhang, R.; Fan, M.; Feng, G.; Sun, J.; et al. Design of catalysts for selective CO2 hydrogenation. *Nat. Synth.* **2025**, *4* (3), 288-302.
(3) Sun, R.; Liao, Y.; Bai, S.-T.; Zheng, M.; Zhou, C.; Zhang, T.; Sels, B. F. Heterogeneous catalysts for CO2 hydrogenation to formic acid/formate: from nanoscale to single atom. *Energy Environ. Sci.* **2021**, *14* (3), 1247-1285.
(4) Kattel, S.; Ramírez, P. J.; Chen, J. G.; Rodriguez, J. A.; Liu, P. Active sites for CO2 hydrogenation to methanol on Cu/ZnO catalysts. *Science* **2017**, *355* (6331), 1296-1299.
(5) Gao, P.; Li, S.; Bu, X.; Dang, S.; Liu, Z.; Wang, H.; Zhong, L.; Qiu, M.; Yang, C.; Cai, J.; et al. Direct conversion of CO2 into liquid fuels with high selectivity over a bifunctional catalyst. *Nat. Chem.* **2017**, *9* (10), 1019-1024.
(6) Beck, A.; Zabilskiy, M.; Newton, M. A.; Safonova, O.; Willinger, M. G.; van Bokhoven, J. A. Following the structure of copper-zinc-alumina across the pressure gap in carbon dioxide hydrogenation. *Nat. Catal.* **2021**, *4* (6), 488-497.
(7) Nishimura, H.; Yatsu, T.; Fujitani, T.; Uchijima, T.; Nakamura, J. Synthesis and decomposition of formate on a Cu(111) surface — kinetic analysis. *J. Mol. Catal. A Chem.* **2000**, *155* (1), 3-11.
(8) Wang, G.; Morikawa, Y.; Matsumoto, T.; Nakamura, J. Why Is Formate Synthesis Insensitive to Copper Surface Structures? *J. Phys. Chem. B* **2006**, *110* (1), 9-11.
(9) Nakano, H.; Nakamura, I.; Fujitani, T.; Nakamura, J. Structure-Dependent Kinetics for Synthesis and Decomposition of Formate Species over Cu(111) and Cu(110) Model Catalysts. *J. Phys. Chem. B* **2001**, *105* (7), 1355-1365.
(10) Quan, J.; Kondo, T.; Wang, G.; Nakamura, J. Energy transfer dynamics of formate decomposition on Cu(110). *Angew. Chem. Int. Ed.* **2017**, *56* (13), 3496-3500.
(11) Muttaqien, F.; Oshima, H.; Hamamoto, Y.; Inagaki, K.; Hamada, I.; Morikawa, Y. Desorption dynamics of CO2 from formate decomposition on Cu(111). *Chem. Commun.* **2017**, *53* (66), 9222-9225.
(12) Yin, R.; Xia, J.; Jiang, B.; Guo, H. Theoretical Insights into Structure Sensitivity in Formate Decomposition Dynamics on Cu Surfaces. *ACS Catal.* **2023**, *13* (21), 14103-14111.
(13) Quan, J.; Muttaqien, F.; Kondo, T.; Kozarashi, T.; Mogi, T.; Imabayashi, T.; Hamamoto, Y.; Inagaki, K.; Hamada, I.; Morikawa, Y.; et al. Vibration-driven reaction of $CO_2$ on Cu surfaces via Eley–Rideal-type mechanism. *Nat. Chem.* **2019**, *11*, 722–729.
(14) Meng, G.; Zhang, Y.; Jiang, B.; Guo, H. Dynamics of Surface Processes: Impact of Adiabatic and Nonadiabatic Energy Dissipation. *Annu. Rev. Phys. Chem.* **2026**, *77*, 61-84.
(15) Zhang, Y.; Hu, C.; Jiang, B. Embedded atom neural network potentials: Efficient and accurate machine learning with a physically inspired representation. *J. Phys. Chem. Lett.* **2019**, *10* (17), 4962-4967.
(16) Kresse, G.; Furthmuller, J. Efficient iterative schemes for ab initio total-energy calculations using plane wave basis set. *Phys. Rev. B* **1996**, *54*, 11169-11186.
(17) Klimeš, J.; Bowler, D. R.; Michaelides, A. Chemical accuracy for the van der Waals density functional. *J. Phys.: Condens. Matter* **2010**, *22* (2), 022201.
(18) Muttaqien, F.; Hamamoto, Y.; Hamada, I.; Inagaki, K.; Shiozawa, Y.; Mukai, K.; Koitaya, T.; Yoshimoto, S.; Yoshinobu, J.; Morikawa, Y. CO2 adsorption on the copper surfaces: van der Waals density functional and TPD studies. *J. Chem. Phys.* **2017**, *147* (9), 094702.
(19) Jiang, Z.; Zhang, L.; Bonnet, L.; Yang, D.; Jiang, B. A practical quasi-classical trajectory method to avoid zero-point energy leakage in dissociative chemisorption of polyatomic molecules on surfaces. *J. Chem. Phys.* **2025**, *163* (1), 014706.
(20) Smith, R. R.; Killelea, D. R.; DelSesto, D. F.; Utz, A. L. Preference for vibrational over translational energy in a gas-surface reaction. *Science* **2004**, *304*, 992-995.
(21) Juurlink, L. B. F.; Killelea, D. R.; Utz, A. L. State-resolved probes of methane dissociation dynamics. *Prog. Surf. Sci.* **2009**, *84* (3), 69-134.
(22) Hundt, P. M.; Jiang, B.; van Reijzen, M.; Guo, H.; Beck, R. D. Vibrationally promoted dissociation of water on Ni(111). *Science* **2014**, *344*, 504-507.
(23) Farjamnia, A.; Jackson, B. The dissociative chemisorption of $CO_2$ on Ni(100): A quantum dynamics study. *J. Chem. Phys.* **2017**, *146* (7), 074704.
(24) Jiang, B.; Guo, H. Communication: Enhanced dissociative chemisorption of $CO_2$ via vibrational excitation. *J. Chem. Phys.* **2016**, *144* (9), 091101.
(25) Yin, R.; Guo, H. Multidimensional Dynamics of CO2 Dissociative Chemisorption on Cu(110). *J. Phys. Chem. C* **2024**, *128* (23), 9483-9491.
(26) Gonzalez, F. J.; Tachino, C. A.; Busnengo, H. F. CO2 dissociative sticking on Cu(110). *J. Chem. Phys.* **2026**, *164* (21), 214701.
(27) Michelsen, H. A.; Rettner, C. T.; Auerbach, D. J.; Zare, R. N. Effect of rotation on the translational and vibrational energy dependence of the dissociative adsorption of $D_2$ on Cu(111). *J. Chem. Phys.* **1993**, *98* (10), 8294-8307.
(28) Rettner, C. T.; Michelsen, H. A.; Auerbach, D. J. Quantum-state-specific dynamics of the dissociative adsorption and associative desorption of $H_2$ at a Cu(111) surface. *J. Chem. Phys.* **1995**, *102* (11), 4625-4641.
(29) Juurlink, L. B. F.; Smith, R. R.; Utz, A. L. The role of rotational excitation in the activated dissociative chemisorption of vibrationally excited methane on Ni(100). *Faraday Disc.* **2000**, *117*, 147-160.
(30) Jiang, B. Rotational and steric effects in water dissociative chemisorption on Ni(111). *Chem. Sci.* **2017**, *8* (9), 6662-6669.
(31) Jiang, B.; Guo, H. Origin of steric effects in methane dissociative chemisorption. *J. Phys. Chem. C* **2016**, *120*, 8220–8226.
(32) Gerrits, N.; Geweke, J.; Auerbach, D. J.; Beck, R. D.; Kroes, G.-J. Highly Efficient Activation of HCl Dissociation on Au(111) via Rotational Preexcitation. *J. Phys. Chem. Lett.* **2021**, *12* (30), 7252-7260.
(33) Liu, T.; Meng, K. Harnessing work-function-driven rotational steering for quantum state control in HCl dissociation on bimetallic alloys. *Chem. Sci.* **2026**, *17* (12), 6187-6196.
(34) Liu, T.; Fu, B.; Zhang, D. H. Highly enhanced reactivity of HCl on the Ag/Au(111) alloy surface via rotational quantum state excitation. *J. Chem. Phys.* **2024**, *161* (23), 234305.
(35) Nakamura, J.; Kushida, Y.; Choi, Y.; Uchijima, T.; Fujitani, T. X-ray photoelectron spectroscopy and scanning tunnel microscope studies of formate species synthesized on Cu(111) surfaces. *J. Vac. Sci. Technol. A* **1997**, *15* (3), 1568-1571.
(36) Guo, H.; Jiang, B. The sudden vector projection model for reactivity: Mode specificity and bond selectivity made simple. *Acc. Chem. Res.* **2014**, *47* (12), 3679-3685.
(37) Jiang, B.; Li, J.; Guo, H. Effects of reactant rotational excitation on reactivity: Perspectives from the sudden limit. *J. Chem. Phys.* **2014**, *140*, 034112.

(38) Levine, R. D. The chemical shape of molecules: an introduction to dynamic stereochemistry. *J. Phys. Chem.* **1990**, *94* (26), 8872-8880.
(39) Wang, J.; Meeprasert, J.; Han, Z.; Wang, H.; Feng, Z.; Tang, C.; Sha, F.; Tang, S.; Li, G.; Pidko, E. A.; et al. Highly dispersed Cd cluster supported on TiO2 as an efficient catalyst for CO2 hydrogenation to methanol. *Chin. J. Catal.* **2022**, *43* (3), 761-770.
(40) Yang, Y.; Han, Y.; Liu, H.; Xie, W.; Hu, P. First-Principles Study on the Activation Mechanism during CO2 Hydrogenation Utilizing the ZnZrO Catalytic System. *J. Phys. Chem. C* **2025**, *129* (33), 14739-14746.

**Supporting Information for**

# Mode-Specific Dynamics of $CO_2$ Hydrogenation on Copper: The Hidden Role of Molecular Rotation

Junfan Xia[1], Zhikai Jiang[1], Yaolong Zhang[1], Bo Peng[2], Hua Guo[3], and Bin Jiang[1,4*]

1. State Key Laboratory of Precision and Intelligent Chemistry, Department of Chemical Physics, University of Science and Technology of China, Hefei, Anhui 230026, China
2. State Key Laboratory of Petroleum Molecular and Process Engineering, SINOPEC Research Institute of Petroleum Processing Co., Ltd., Beijing 100083, China.
3. Department of Chemistry and Chemical Biology, Center for Computational Chemistry, University of New Mexico, Albuquerque, New Mexico 87131, USA
4. Hefei National Laboratory, University of Science and Technology of China, Hefei, 230088, China

*Corresponding author: bjiangch@ustc.edu.cn

# 1. Methods

## 1.1 Density functional theory calculations

Total energy calculations and geometric optimizations were performed using Density functional theory (DFT) as implemented in the Vienna Ab initio Simulation Package (VASP)[1]. The non-local optPBE-vdW functional[2] was employed to account for van der Waals (vdW) interactions. Exchange–correlation effects were treated within the generalized gradient approximation (GGA)[3], and the interaction between ionic cores and valence electrons was described using the projector augmented wave (PAW) method.[4] The plane-wave basis set was truncated at kinetic energies of 490 eV and 5440 eV for the wavefunction and the augmented charge density, respectively. The Brillouin zone was sampled using a 4 × 4 × 1 Monkhorst–Pack k-point mesh[5]. A Fermi smearing of 0.1 eV was applied to extrapolate the total energy to the zero-temperature limit.

The optimized lattice parameter for bulk Cu is 3.647 Å, agreeing well with the experimental value of 3.615 Å.[6] The Cu(111) surface was modelled by a four-layer slab in a 4×4 supercell, with a vacuum space of 17.5 Å between periodic slabs and a dipole correction in the Z direction. The top three layers of Cu atoms were moveable and the bottom layer was fixed. Eight hydrogen atoms were pre-adsorbed on both fcc and hcp threefold hollow sites of the surface as in Ref. 7, in alignment with the experimental coverage of 0.5 ML and the (2×2) pattern in low energy electron diffraction (LEED) experiment[8].

### 1.2 Embedded Atom neural network potential

To describe the reaction between $CO_2$ on a H-adsorbed Cu(111) facet, the potential energy surface (PES) was constructed by means of the Embedded Atom Neural Network (EANN) approach[9, 10]. In this atomistic framework, the total energy of the system is regarded as the sum of atomic energies, each of which is an output of an element-wise NN determined by the electron density of this atom embedded in the environment contributed by other atoms nearby,

$$E = \sum_{i=1}^{N} E_i = \sum_{i=1}^{N} \mathrm{NN}_i\left(\rho_i\right). \tag{1}$$

For simplicity, the embedded atom density (EAD) descriptor ($\rho_i$) is expanded by Gauss-type orbitals (GTOs) centered at its neighboring atoms, consisting of multiple orbital-dependent density features,

$$\rho_i = \sum_{l_x,l_y,l_z}^{l_x+l_y+l_z=L} \frac{L!}{l_x!l_y!l_z!}\left(\sum_{j=1}^{n_{\text{atom}}} c_j \varphi_{l_x l_y l_z}^{\alpha,r_s}(\mathbf{r}_{ij}) f_c\left(\mathbf{r}_{ij}\right)\right)^2, \tag{2}$$

where $n_{\text{atom}}$ is the number of atoms lying nearby the embedded atom within a cutoff radius $r_c$ and $f_c(r_{ij})$ is a cutoff function to ensure that the contribution of each neighbor atom decays smoothly to zero at $r_c$, $c_j$ is an element-dependent expansion coefficient of an atomic orbital at atom $j$, which is adjustable together with the element dependent NN parameters. The GTO is expressed in the Cartesian coordinate frame,

$$\varphi_{l_x l_y l_z}^{\alpha,r_s}(\mathbf{r}_{ij}) = x_{ij}^{l_x} y_{ij}^{l_y} z_{ij}^{l_z} \exp\left(-\alpha\left|r_{ij} - r_s\right|^2\right), \tag{3}$$

where $\mathbf{r}_{ij} = (x_{ij}, y_{ij}, z_{ij})$ is the Cartesian coordinate vector of the embedded atom $i$ relative to atom $j$, $r_{ij}$ is the internuclear distance, $L$ is the total angular momentum, with $l_x$, $l_y$, and $l_z$ being its projections onto each axis, respectively, collectively determining the angular shape of the GTO, while $\alpha$ and $r_s$ determine its radial distribution. In this system, these hyperparameters $\alpha$ and $r_s$ of GTOs were selected ensure linear independence with respect to one another[11], with the maximum value of $L$ set to 3 and $r_c$ =5.6 Å, yielding a total of 24 EAD features. Each atomic NN has two hidden layers with 40 and 60 neurons, respectively.

An uncertainty-driven active learning scheme[12] was applied to construct the training dataset. In this scheme, a weighted negative of squared difference surface (WNSDS) is defined by two trial EANN PESs with different parameters,

$$F(\mathbf{x}) = -\omega(y_1, y_2)(y_1(\mathbf{x}) - y_2(\mathbf{x}))^2, \quad (4)$$

in which $\mathbf{x}$ is the collection of all coordinates of configurations, $y_1$ and $y_2$ are the two NN output energies used in defining the WNSDS. By definition, WNSDS is small at existing data locations, while largely negative in regions that are distant from existing samples. The weighting function $\omega(y_1, y_2)$ depends exponentially on the average energy of the two trial PESs—a higher weight is assigned for a lower-energy region. This controlling factor avoids over-sampling the highly distorted configurations deviating from the dynamically important regions with high energies.

In practice, starting from two preliminary EANN PESs trained on a set of seed points along the minimum energy path (MEP), we searched for local minima on the initial WNSDS. These minima correspond to high uncertainty locations on the PES, where *ab initio* calculations were performed and the resulting data were added into the training set. The updated training set was then used to refine the trial PESs and the WNSDS was recomputed to guide the next iteration of sampling. In this iterative way, the dataset was expanded to eventually contain 11000 data points in total, including potential energies and atomic forces. These data points were randomly divided into a training set (90%) and a test set (10%). The training set was used to optimize the model parameters, while the test set served to validate the final EANN PES.

The accuracy of resultant EANN PES has been validated in several aspects. First, the root-mean-square-errors (RMSEs) of these data in the test set are 0.54 meV/atom for energy and 33.4 meV/Å for atomic force, respectively. Second, the PES reproduces the DFT-derived phonon band structure and density of phonon states of the Cu(111) slab quite well, as evidenced in Figure S1. This indicates an adequate description for the motion of surface atoms. Third, the PES reproduces well the stationary structures and energies along the MEP, as shown in Figure 1 and Table S1.

**1.3 Quasi-classical trajectory calculations**

All quasi-classical trajectory (QCT) calculations were performed on the EANN PES with our in-house heavily-modified VENUS code[13]. Specifically, for simulating the molecular beam conditions in Ref. 14, the vibrational and rotational states of $CO_2$ were sampled from Boltzmann distributions at corresponding vibrational ($T_v$) and rotational ($T_r$) temperatures. In typical supersonic molecular beams, rotational cooling is often quite efficient, while vibrational temperature remains hot.[14] Therefore, $T_v$ was set equal to the nozzle temperature ($T_n$), and $T_r$ were taken as 5% of $T_n$ (except in one special case where $T_r$ = 0 K for comparison). For initial state-selected calculations, the initial vibrational state of $CO_2$ was sampled with the standard normal mode sampling scheme.[15] The linear $CO_2$ molecule is treated as a symmetric top with a rotational quantum number ($J$) and its projection onto the molecular axis ($K$=0). The rotational angular momenta were sampled as $\boldsymbol{j}=\sqrt{J(J+1)}\hbar$, $j_z = K\hbar$, $j_x=(j^2-j_z^2)^{1/2}\sin(2\pi R)$, and $j_y=(j^2-j_z^2)^{1/2}\cos(2\pi R)$, where $R$ is a random number.[16] After the initial sampling of positions and momenta for a given ro-vibrational state, the $CO_2$ molecule was randomly oriented and placed 8 Å above the Cu(111) surface, with its lateral position randomized within the surface unit cell. Its incident velocity was assigned according to the mean incidence energy along surface normal toward the surface in experiments[14]. The initial surface configurations were randomly sampled from the snapshots of equilibrated molecular dynamics simulations of the bare surface at an experimental surface temperature ($T_s$).

All trajectories were propagated using the velocity Verlet algorithm, with a maximum propagation time of 20 ps and a time step of 0.1 fs, conserving the total energy within ~0.1 meV on average. A trajectory was considered as "reactive" if an HCOO species was formed. Since the stable HCOO$^*$ adsorbate features a C-H distance of ~1.1 Å, the dynamic formation of an HCOO species during trajectories was identified when $CO_2$ connects any surface H atom with a C-H distance of less than 1.2 Å persisted for at least 100 fs. Varying this threshold within a range of 1.2 to 1.4 Å was found to have no noticeable influence on the results, indicating its robustness. Otherwise, a trajectory was classified as "non-reactive", including cases where $CO_2$ scatters away from the surface or becomes trapped in the physisorption well without reaction up to the maximum propagation time. Neither COOH$^*$ formation nor $CO_2$ dissociation was observed in our QCT calculations.

### 1.4 Product state analysis and adsorbate Gaussian binning

The reaction probability ($P_0$) of QCT calculations can be easily obtained as the ratio of the number of "reactive" trajectories ($N_r$) to the total number of trajectories ($N_{tot}$), *i.e.*, $P_0=N_r/N_{tot}$. However, it was recently found that this treatment may significantly overestimate the reactivity in certain systems—for example, dissociative chemisorption on surfaces—especially when there is substantial zero-point energy leakage into the reaction coordinate and the reactivity is low. In such cases, performing a normal-mode analysis of the vibrational states[17] for both scattered and adsorbed products, followed by Gaussian binning of the resulting state populations, has been found to largely correct the QCT-predicted reaction probability, bringing it into close agreement with quantum mechanical and experimental results under the same conditions.[18, 19]. This so-called adsorbate Gaussian binning (AGB) approach is adopted in this work to calculate the reaction probability of QCT calculations.

The procedure of the QCT-AGB approach is described in the following. First, for a given polyatomic product, the translational and rotational momenta are removed, leaving only the vibrational momenta. To obtain the classical vibrational action number $(n_i)$ for each normal mode, normal mode analysis[17] is performed. The mass-weighted coordinates of the product $\mathbf{q}$ is aligned with the corresponding reference configuration $\mathbf{q}_{\mathrm{ref}}$ using a quaternion-based method[20]. The displacement along each normal mode and its conjugate momentum are then extracted from the aligned coordinates to calculate $n_i$. Specifically, for scattered molecules, $\mathbf{q}_{\mathrm{ref}}$ corresponds to the gas-phase equilibrium geometry of the molecule, and for adsorbates on surface, $\mathbf{q}_{\text{ref}}$ is taken as the most stable adsorption configuration on surface. We have shown that this normal mode analysis is effective for both, as their internal vibrational modes are largely decoupled from other degrees of freedom[18, 19]. Next, vibrational quantization is performed using the energy-based Gaussian binning (1GB) algorithm,

$$G(\boldsymbol{n})=\frac{1}{\sqrt{\pi\varepsilon}}\exp\left\{-\left[\frac{E(\boldsymbol{n}')-E(\boldsymbol{n})}{2\varepsilon E(\mathbf{0})}\right]^2\right\}. \tag{5}$$

In this expression, $G(\boldsymbol{n})$ is the Gaussian weight of the trajectory, the vibrational state of the molecule ($\boldsymbol{n}$) is labeled as a set of vibrational quantum numbers $\{n_i\}$, $\boldsymbol{n}'$ represents the corresponding set of vibrational action numbers $\{n_i'\}$. The full-width at half-maximum of the Gaussian function ($\lambda$) is 0.03 in this work and $\varepsilon=\lambda/2(\ln 2)^{1/2}$.

$E(\mathbf{0})$ is the ZPE for the reference configuration, $E(\boldsymbol{n}')=\sum_{i=1}^{3N-5}\omega_i\left(n'+\frac{1}{2}\right)$ and $E(\boldsymbol{n})=\sum_{i=1}^{3N-5}\omega_i\left(n+\frac{1}{2}\right)$ are the actual vibrational energy given by classical fractional action numbers and the harmonic vibrational energy given by integer action numbers, where their energy difference determines the Gaussian weight. Finally, one can estimate the GB-corrected reaction probability by,

$$P_0=\frac{\sum_{i=1}^{N_{\text{tot}}}\delta_i G_i(\boldsymbol{n})}{\sum_{i=1}^{N_{\text{tot}}}G_i(\boldsymbol{n})} \tag{6}$$

When $i$th trajectory is reactive $\delta_i=1$, otherwise $\delta_i=0$. Additionally, for the AGB method, since each trajectory is assigned an independent weight, we use the variance formula as follows[21],

$$\sigma = \sqrt{\frac{\sum_{i=1}^{N_{\text{tot}}} G_i^2 P_0 \left(1 - P_0\right)}{\left(\sum_{i=1}^{N_{\text{tot}}} G_i\right)^2}} \tag{7}$$

**1.5 S-shaped reactivity curve**

Following the experimental work[14], we fit our QCT data of the initial reaction probability ($P_0$) to S-shaped reactivity curves as a function of the translational energy ($E_t$) and mean vibrational energy ($\bar{E}_v$). This fitting is based on the following function proposed by Luntz[22],

$$P_0(E) = \frac{A}{2} \times \left(1 + \text{erf}\left|\frac{E - E_0}{W}\right|\right), \tag{8}$$

which assumes that the barrier height follows a Gaussian distribution with a width of $W$. In this expression, $A$ and $E_0$ represent the asymptotic value of $P_0$ at high energy and the energy at which $P_0$ reaches half of its asymptotic value, erf is the error function. Optimal parameters for fitting S-shaped curves in Figures 2 and 3 are summarized in Table S3.

**1.6 Sudden Vector Projection**

The Sudden Vector Projection (SVP) model has proven to reasonably predict the mode specificity and bond selectivity in gas phase and gas-surface reactions.[23, 24] It is based on the hypothesis that the timescale of the collision is short enough (sudden limit) for any internal energy redistribution within the reactant, which is applicable to this direct Eley-Rideal reaction. An initial state with a large coupling with the reaction coordinate at the transition state is thus expected to give rise to a large vibrational efficacy. In the classical normal mode picture, this coupling can be quantified by the alignment between the corresponding normal mode vectors. In practice, the molecular structure is optimized in the asymptote followed by a normal mode analysis to obtain the reactant vectors. The same procedure is done at the transition state whose structure is reoriented to have a maximal overlap with the reactant. The SVP values are then computed by projecting a reactant normal mode (or translational) vector $\boldsymbol{Q}_i$ onto the reaction coordinate $\boldsymbol{Q}_{\mathrm{RC}}$ of the transition state corresponding to the imaginary frequency, namely $p_i = \boldsymbol{Q}_i \cdot \boldsymbol{Q}_{\mathrm{RC}}$. The SVP model is essentially an extension of Polanyi's rules[25] to a multidimensional PES that is applicable to polyatomic reactions.

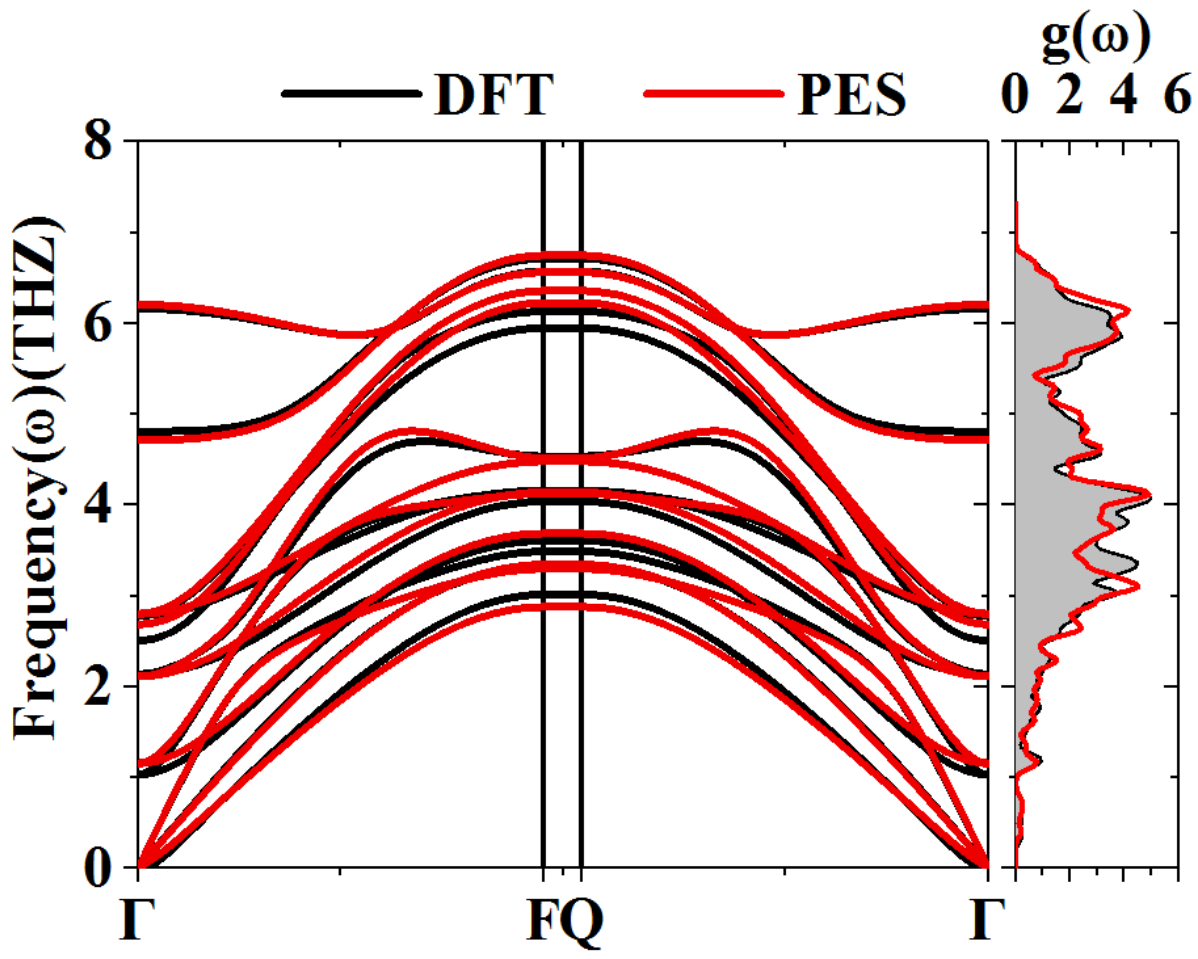

**Figure S1**. Phonon dispersion of Cu(111) obtained by DFT and EANN PES along the path through the surface Brillouin zone given by the high-symmetry points Γ-F-Q-Γ (left) and corresponding phonon density of states as a function of the frequency of phonon modes (right).

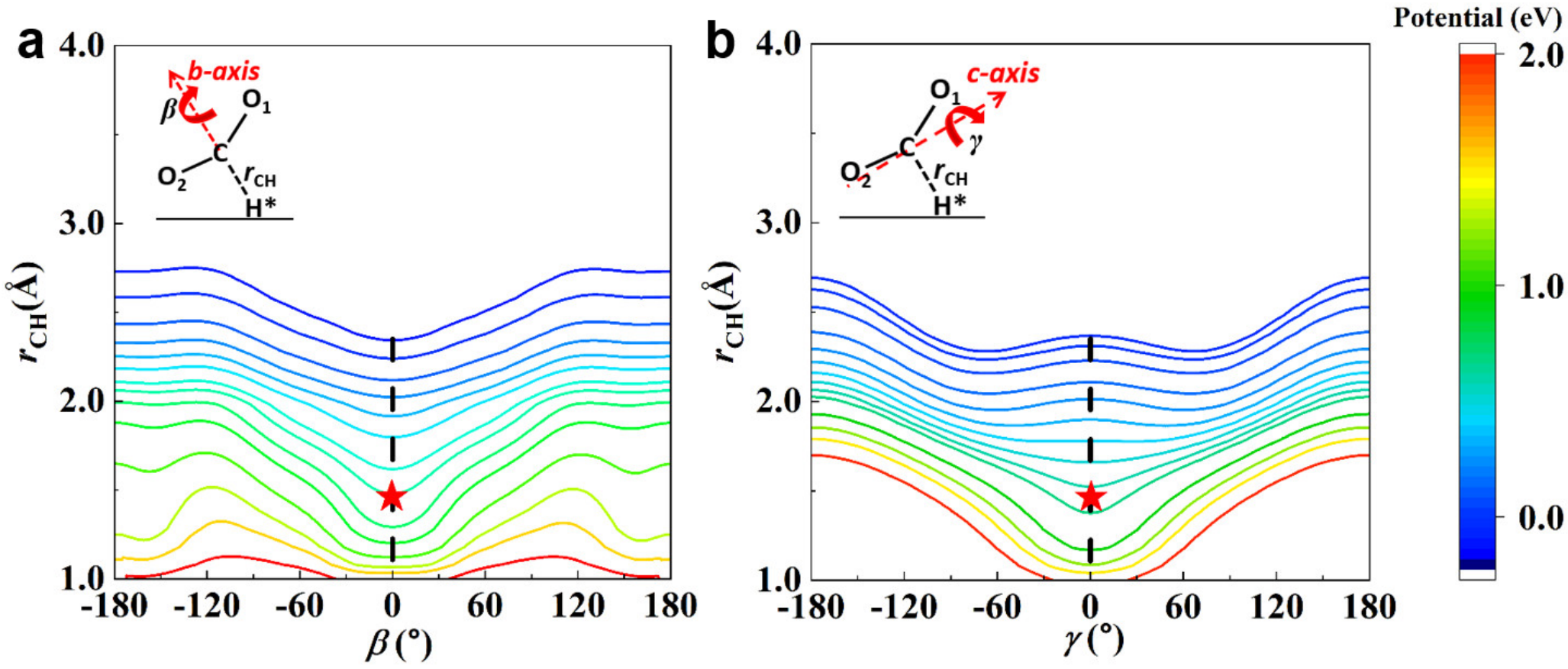

**Figure S2**. Two-dimensional cuts of PESs as a function of the C-H distance ($r_{\mathrm{CH}}$) and the orientation angle (a) $\beta$ and (b) $\gamma$ with respect to rotations about b and c axis defined in the main text, with the H positions and the molecular approaching direction fixed at their values of TS. The red stars indicate the TS for formate formation.

**Table S1**. Comparison of the structure parameters of initial states ($CO_2(g)+H^*$), physisorption ($CO_2^*+H^*$), Transition States (TS), monodentate formate (m-$HCOO^*$) and bidentate formate (bi-$HCOO^*$) obtained via EANN PES and DFT. Energies and structure parameters in parentheses are obtained from DFT for comparison. $\theta$ is the O-C-O angle.

| | **C-$O_1$ (Å)** | **C-$O_2$ (Å)** | **C-H (Å)** | **$\theta$ (deg)** | **$E$ (eV)** |
|---|---|---|---|---|---|
| $H^*+CO_2$ (g) | 1.177<br>(1.177) | 1.178<br>(1.178) | 5.992<br>(5.992) | 180.0<br>(180.0) | 0.000<br>(0.000) |
| $H^*+CO_2^*$ | 1.176<br>(1.177) | 1.179<br>(1.179) | 3.205<br>(3.204) | 180.0<br>(179.9) | -0.223<br>(-0.220) |
| TS | 1.199<br>(1.199) | 1.241<br>(1.241) | 1.480<br>(1.481) | 144.4<br>(144.5) | 0.598<br>(0.599) |
| m-$HCOO^*$ | 1.216<br>(1.219) | 1.352<br>(1.350) | 1.109<br>(1.110) | 124.1<br>(123.8) | 0.056<br>(0.060) |
| bi-$HCOO^*$ | 1.271<br>(1.272) | 1.277<br>(1.279) | 1.109<br>(1.109) | 126.8<br>(126.8) | -0.441<br>(-0.436) |

**Table S2**. Final vibrational state distributions of scattered $CO_2$ molecules from the incidence of the rotational ground state ($J$=0) and excited state ($J$=10, values in parentheses), both in the vibrational ground state, at various translational energies.

| $E_t$(eV) | Final Vibrational State Populations | | |
|---|---|---|---|
| | **(000)** | **(010)** | **(020)** |
| 2.43 | 0.625 (0.553) | 0.109 (0.146) | 0.047 (0.058) |
| 2.20 | 0.700 (0.595) | 0.098 (0.145) | 0.036 (0.057) |
| 1.97 | 0.776 (0.686) | 0.078 (0.117) | 0.027 (0.045) |
| 1.82 | 0.821 (0.751) | 0.063 (0.106) | 0.022 (0.039) |
| 1.68 | 0.863 (0.783) | 0.053 (0.097) | 0.017 (0.034) |

**Table S3**. Parameters for fitting the S-shaped curves based on the calculated $P_0$ data at various $T_s$, $E_t$, $\bar{E}_v$ in Figures 2 and 3.

| **Condition** | **Dependence** | **A** | $E_0$ **(eV)** | **W (eV)** |
|---|---|---|---|---|
| $T_r$=0 | $T_s$ | 0.0021 | 1.0000 | 1.0000 |
| $T_r$=0 | $E_t$ | 0.0038 | 1.9767 | 0.6225 |
| $T_r$=0 | $\bar{E}_v$ | 0.0003 | 0.2200 | 0.1080 |
| $T_r$=5%$T_n$ | $\bar{E}_v$ | 0.0024 | 0.2247 | 0.0900 |
| (000,0) | $E_t$ | 0.0011 | 2.3173 | 0.3428 |
| (100,0) | $E_t$ | 0.0039 | 2.2595 | 0.3572 |
| (010,0) | $E_t$ | 0.0024 | 2.1962 | 0.3801 |
| (001,0) | $E_t$ | 0.0036 | 2.2790 | 0.3675 |
| (000,10) | $E_t$ | 0.0147 | 2.7000 | 0.6769 |

## References

(1) Kresse, G.; Furthmuller, J. Efficient iterative schemes for ab initio total-energy calculations using plane wave basis set. *Phys. Rev. B* **1996**, *54*, 11169-11186.
(2) Klimeš, J.; Bowler, D. R.; Michaelides, A. Chemical accuracy for the van der Waals density functional. *J. Phys.: Condens. Matter* **2010**, *22* (2), 022201.
(3) Perdew, J. P.; Burke, K.; Ernzerhof, M. Generalized gradient approximation made simple. *Phys. Rev. Lett.* **1996**, *77*, 3865-3868.
(4) Blöchl, P. E. Projector augmented-wave method. *Phys. Rev. B* **1994**, *50*, 17953-17979.
(5) Monkhorst, H. J.; Pack, J. D. Special points for Brillouin-zone integrations. *Phys. Rev. B* **1976**, *13*, 5188-5192.
(6) Straumanis, M. E.; Yu, L. S. Lattice parameters, densities, expansion coefficients and perfection of structure of Cu and of Cu-In [alpha] phase. *Acta Crystallogr. A* **1969**, *25* (6), 676-682.
(7) Luo, M. F.; Hu, G. R.; Lee, M. H. Surface structures of atomic hydrogen adsorbed on Cu(111) surface studied by density-functional-theory calculations. *Surf. Sci.* **2007**, *601* (6), 1461-1466.
(8) McCash, E. M.; Parker, S. F.; Pritchard, J.; Chesters, M. A. The adsorption of atomic hydrogen on Cu(111) investigated by reflection-absorption infrared spectroscopy, electron energy loss spectroscopy and low energy electron diffraction. *Surf. Sci.* **1989**, *215* (3), 363-377.
(9) Zhang, Y.; Hu, C.; Jiang, B. Embedded atom neural network potentials: Efficient and accurate machine learning with a physically inspired representation. *J. Phys. Chem. Lett.* **2019**, *10* (17), 4962-4967.
(10) Zhang, Y.; Xia, J.; Zhang, Y.; Jiang, B. REANN 2.0: An Efficient Package of Neural Network Potentials for Multi-Element Systems. *Chin. J. Chem. Phys.* **2025**, *38* (6), 797-806.
(11) Xia, J.; Zhang, Y.; Jiang, B. Efficient Selection of Linearly Independent Atomic Features for Accurate Machine Learning Potentials. *Chin. J. Chem. Phys.* **2021**, *34* (6), 695-703.
(12) Lin, Q.; Zhang, L.; Zhang, Y.; Jiang, B. Searching Configurations in Uncertainty Space: Active Learning of High-Dimensional Neural Network Reactive Potentials. *J Chem. Theory Comput.* **2021**, *17* (5), 2691-2701.
(13) Hu, X.; Hase, W. L.; Pirraglia, T. Vectorization of the general Monte Carlo classical trajectory program VENUS. *J. Comput. Chem.* **1991**, *12*, 1014-1024.
(14) Quan, J.; Muttaqien, F.; Kondo, T.; Kozarashi, T.; Mogi, T.; Imabayashi, T.; Hamamoto, Y.; Inagaki, K.; Hamada, I.; Morikawa, Y.; et al. Vibration-driven reaction of $CO_2$ on Cu surfaces via Eley–Rideal-type mechanism. *Nat. Chem.* **2019**, *11*, 722–729.
(15) Hase, W. L. Classical trajectory simulations: Initial conditions. In *Encyclopedia of Computational Chemistry*, Alinger, N. L. Ed.; Vol. 1; Wiley, 1998; pp 399-402.
(16) Lourderaj, U.; Martínez-Núñez, E.; Hase, W. L. Representing and Selecting Vibrational Angular Momentum States for Quasiclassical Trajectory Chemical Dynamics Simulations. *J. Phys. Chem. A* **2007**, *111* (41), 10292-10301.
(17) Corchado, J. C.; Espinosa-Garcia, J. Product vibrational distributions in polyatomic species based on quasiclassical trajectory calculations. *Phys. Chem. Chem. Phys.* **2009**, *11* (43), 10157-10164.
(18) Jiang, Z.; Zhang, L.; Bonnet, L.; Yang, D.; Jiang, B. A practical quasi-classical trajectory method to avoid zero-point energy leakage in dissociative chemisorption of polyatomic molecules on surfaces. *J. Chem. Phys.* **2025**, *163* (1), 014706.
(19) Jiang, Z.; Jiang, B. Quasi-Classical Trajectory with Adsorbate Gaussian Binning: Quantum-State-

Resolved Prediction of Dissociative Sticking Probability Made Simple. *J. Phys. Chem. Lett.* **2026**, *17* (5), 1513-1518.
(20) Horn, B. K. P. Closed-form solution of absolute orientation using unit quaternions. *J. Opt. Soc. Am. A* **1987**, *4* (4), 629-642.
(21) Kish, L. Survey sampling. New York: John Wesley & Sons. *Am. Polit. Sci. Rev.* **1965**, *59* (4), 1025.
(22) Luntz, A. C. A simple model for associative desorption and dissociative chemisorption. *J. Chem. Phys.* **2000**, *113*, 6901-6905.
(23) Guo, H.; Jiang, B. The sudden vector projection model for reactivity: Mode specificity and bond selectivity made simple. *Acc. Chem. Res.* **2014**, *47* (12), 3679-3685.
(24) Chen, J.; Zhou, X.; Zhang, Y.; Jiang, B. Vibrational control of selective bond cleavage in dissociative chemisorption of methanol on Cu(111). *Nat. Commun.* **2018**, *9* (1), 4039.
(25) Polanyi, J. C. Some concepts in reaction dynamics. *Science* **1987**, *236*, 680-690.